\documentclass[twocolumn,prb,citeautoscript,superscriptaddress]{revtex4}
\usepackage{amsfonts,amsmath,amssymb}
\usepackage[dvips]{graphicx,color}
\input epsf 
\newcommand{\dir}{Figs}

\newcommand{\rr}{ {\mathbf{r}} }
\newcommand{\hr}{ {\mathbf{\hat{r}}} }
\newcommand{\uu}{ {\mathbf{\hat{u}}} }

\newcommand{\qq}{ {\mathbf{q}} }
\newcommand{\Ptens}{\mbox{\boldmath $\mathsf{P}$}}
\newcommand{\Qtens}{\mbox{\boldmath $\mathsf{Q}$}}
\newcommand{\Itens}{\mbox{\boldmath $\mathsf{I}$}}
\newcommand{\sigtens}{\boldsymbol{\sigma}}
\newcommand{\ie}{{\em i.~e.}, }
\newcommand{\eg}{{\em e.~g.}, }

\begin{document}

\title{Nematic-Isotropic Interfaces under Shear: A Molecular Dynamics
Simulation.}

\author{Guido Germano}
\email{germano@staff.uni-marburg.de}
\affiliation{Fakult\"at f\"ur Physik, Universit\"at Bielefeld,
Universit\"atsstra{\ss}e 25, D-33615 Bielefeld, Germany}
\affiliation{
Fachbereich Chemie, Philipps-Universit\"at Marburg,
D-35032 Marburg, Germany}

\author{Friederike Schmid}
\email{schmid@physik.uni-bielefeld.de}
\affiliation{Fakult\"at f\"ur Physik, Universit\"at Bielefeld,
Universit\"atsstra{\ss}e 25, D-33615 Bielefeld, Germany}

\date{\today}
\begin{abstract}
\noindent

We present a large-scale molecular dynamics study of 
nematic-paranematic interfaces under shear. 
We use a model of soft repulsive ellipsoidal particles
with well-known equilibrium properties, and consider
interfaces which are oriented normal to the direction
of the shear gradient (common stress case). The director at the 
interface is oriented parallel to the interface (planar).
A fixed average shear rate is imposed with Lees-Edwards 
boundary conditions, and the heat is dissipated with a 
profile-unbiased thermostat. First we study the properties 
of the interface at one particular shear rate in detail. 
The local interfacial profiles and the capillary wave 
fluctuations of the interfaces are calculated and compared 
with those of the corresponding equilibrium interface. 
Under shear, the interfacial width broadens and the 
capillary wave amplitudes at large wavelengths increase.
The strain is distributed inhomogeneously in the system 
(shear banding), the local shear rate in the nematic region 
being distinctly higher than in the paranematic region. 
Surprisingly, we also observe (symmetry breaking) flow 
in the {\em vorticity} direction, with opposite direction 
in the nematic and the paranematic state. 
Finally, we investigate the stability of the interface 
for other shear rates and construct a nonequilibrium 
phase diagram.

\end{abstract}
\maketitle

\section{Introduction}
\label{introduction}

Complex fluids often show peculiar behavior under 
shear~\cite{larson_book}. Particularly remarkable phenomena 
are observed in fluids with anisotropic components, which 
can exhibit local or global orientational 
order~\cite{degennes_book,rey_02a}. 
Shear flow tends to orient the materials, which usually 
reduces the shear viscosity 
(shear thinning)~\cite{doi_book, morriss_91a,yuan_97a,
forster_05a}, but can also have the opposite 
effect~\cite{morriss_91a,berret_02a}. In the vicinity of
equilibrium phase transitions, shear flow may shift the
boundaries and enforce ordering transitions~\cite{olmsted_90a,
berret_94a,berret_94b,cappelaere_97a, berret_98a,mather_97a}. 
It may even induce the formation of new structures
that are inhomogeneous in space or time, such as 
multilamellar vesicle phases in surfactant 
systems~\cite{roux_93a,roux_93b} or tumbling states in 
liquid crystalline polymers~\cite{larson_93a,berret_95a,
zakharov_03a,hess_04a,lettinga_04a}. 

In recent years special attention has been given
to situations where phases ``coexist'' under shear. 
The interest was originally raised by the ``spurt effect'' 
in polymer melts, \ie the observation that the flow rate 
in a pipe may change discontinuously as a function of the 
applied pressure difference~\cite{bagley_58a}. 
This was attributed to a nonequilibrium phase transition 
from an entangled polymer phase with high viscosity 
to a highly aligned phase with low 
viscosity~\cite{doi_book,mcleish_86a,cates_93a}. 
The instability results from a nonmonotonicity
in the (hypothetical) homogeneous stress-strain flow 
curve: In a homogeneous system, the stress would
first grow with growing strain rate and then decrease
beyond a certain critical value. Since such a flow curve
is mechanically unstable, the system becomes inhomogeneous. 
The stress plateau in the stress-strain flow curve
was interpreted as the signature of a region where 
phases with different shear rates coexist.
The separation of a fluid into two states with 
distinctly different shear rates is commonly 
called shear banding. 

\begin{figure}[t]
\begin{center}
\includegraphics[width=.4\textwidth]{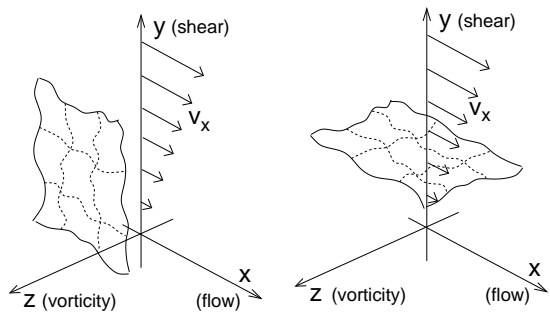}
\caption{Possible orientations of an interface: 
common strain (left) and common stress (right) geometries.
Also indicated is the reference system used throughout this
paper. In this paper, only the common stress case is
investigated.
}
\label{fig:geometries}
\end{center}
\end{figure}

Meanwhile, shear banding has been observed and even 
directly visualized in various systems, most notably 
wormlike micelles~\cite{roux_95a,berret_96a,
britton_99a,fischer_01a,lopez_04a} and rodlike 
viruses~\cite{lettinga_04a}. More generally,
two qualitatively different types of phase 
separation are possible in shear flow~\cite{olmsted_99b}.
These are illustrated in Fig.~\ref{fig:geometries} 
for the Couette geometry. In the shear banding case
described above, the fluid separates in the direction 
of the flow gradient, the phases are subject to the 
same stress, and the shear rates differ (common stress 
case).  Alternatively, the fluid can also phase separate
in the direction perpendicular to both flow 
and flow gradient, the vorticity direction. 
The two phases then share the same shear profile, 
but the stress differs (common strain case). 

The phase separation mechanism described above is 
driven by mechanical (hydrodynamical) forces. 
However, phase separation and shear banding can 
also occur as a result of a thermodynamic phase 
transition -- either a real transition that will 
also occur in the absence of shear flow, or a 
transition between phases that are metastable
(``hidden'') in equilibrium.
Shear banding is often associated with ordering 
phenomena and sometimes occurs in the vicinity of 
a thermodynamic phase boundary. Thermodynamic
forces then contribute to the nonmonotonic
stress-strain relationship which gives rise to
the mechanical instability. Thus the mechanical
and the thermodynamical phase separation cannot always 
be separated unambiguously~\cite{dhont_99a}. 
In fact, theoretical treatments of shear banding 
are often based on free energy expressions
for phase separating systems~\cite{olmsted_92a,
porte_97a, olmsted_97a,olmsted_99a}.

One should, however, note an important difference
between mechanically and thermodynamically
driven phase separation~\cite{schmitt_95a}. 
The signature of mechanically driven phase coexistence 
is the existence of a plateau in the stress-strain 
flow curve. In the common stress case, the
plateau is horizontal, in the common shear case,
it is vertical~\cite{olmsted_99b}.
If one increases the average shear 
rate/average stress in the plateau region, \ie 
the coexistence region, the width of the coexisting 
bands readjusts, but the local stress and the 
shear rate in each phase remain the same. 
For thermodynamically driven phase separation,
however, a free readjustment is only possible if 
solely {\em nonconserved} quantities are 
involved in the phase transition. If the two
coexisting phases differ with respect to the
density of a {\em conserved} quantity (\eg a
concentration), the overall average of this
density imposes an additional constraint. 
As a result, both the band width 
and the coexisting phases must be readjusted, 
and the ``plateau'' acquires a finite 
slope~\cite{olmsted_97a,olmsted_99a,
fielding_03a}.

This raises an important question: How does the 
system determine the location of the plateau and 
select the coexisting states? According to Olmsted 
and coworkers, the selection criterion is the requirement 
that a steady interface between stable coexisting 
homogeneous states exists~\cite{olmsted_92a,olmsted_97a,olmsted_99a}. 
Hence the understanding of the interface provides 
the key to the understanding of the properties of 
the material under shear.

In the present work, we present a molecular dynamics 
simulation of a nematic/isotropic (more precisely, paranematic)
interface under shear, in a situation where 
the interface is already present in the absence 
of shear. The phase separation is thus driven by 
thermodynamic forces. 
The equilibrium properties of isotropic/nematic interfaces 
have been studied by computer simulations in a number
of systems~\cite{bates_97a, allen_00a,albarwani_00a, 
mcdonald_01a,akino_01a,vink_05a,vink_05b}. Our study is
designed such that we can compare the equilibrium
and the nonequilibrium structure of the interface
in as much detail as possible. Therefore, we choose a model 
of soft ellipsoids, which we have investigated extensively 
in the past~\cite{phuong_01a, phuong_03a,lange_02a,
lange_02b, mcdonald_01a,akino_01a}, and used for
large scale simulations of equilibrium 
interfaces~\cite{mcdonald_01a,akino_01a}. These
interfaces are now subjected to shear.
Previous particle-based simulations of liquid 
crystals under shear have focussed on the structure 
and the response functions in homogeneous 
phases~\cite{baalss_86a,sarman_93a,sarman_95a,
sarman_97a,sarman_98a,hess_97a, yuan_97a, bennett_99a,
mcwhirter_02a,mcwhirter_02b,jose_04a}. 
To our best knowledge, our simulation is the 
first particle-based nonequilibrium 
simulation of a liquid crystal interface under flow. 

Our paper is organized as follows: In the
next section, we introduce the model and 
discuss the simulation method. The results
are presented in section~\ref{sec:results}.
We first discuss the stability of the
interface with increasing shear rate, and
then consider in detail the interfacial
structures at one particular shear rate. 
We summarize and conclude in 
section~\ref{sec:conclusions}.

\section{Model and Method}
\label{model}

We consider a system of soft repulsive ellipsoids with length to
width ratio $\sigma_l/\sigma_0 = 15$. Every ellipsoid is characterized
by the position of the center of mass $\rr$ and a unit vector $\uu$ 
pointing in the direction of the long axis. The interaction potential 
between two ellipsoids $i$ and $j$ separated by the center-center
vector $\rr_{ij}$ is given by 
\begin{equation}
U = \left\{ \begin{array}{ll}
4 \epsilon_0 \: (s_{ij}^{-12} - s_{ij}^{-6}) + \epsilon_0,
& \quad s_{ij} < 2^{1/6} \\ 0, & \quad \text{otherwise} \end{array} \right. ,
\end{equation}
where $s_{ij} = (r_{ij}-\sigma(\hr_{ij},\uu_i,\uu_j)+\sigma_0)/\sigma_0$ 
is a scaled and shifted distance and $\sigma$ approximates the contact
distance of two ellipsoids~\cite{berne_72a},
\begin{eqnarray}
\sigma(\hr_{ij},\uu_i,\uu_j) &=& \sigma_0 \Big\{ 1 
- \frac{\chi}{2} \Big[ 
  \frac{ (\uu_i \cdot \hr_{ij} + \uu_j \cdot \hr_{ij})^2}{1+\chi \uu_i \cdot \uu_j}
\\
\nonumber &&  \qquad 
+ \frac{ (\uu_i \cdot \hr_{ij} - \uu_j \cdot \hr_{ij})^2}{1-\chi \uu_i \cdot \uu_j} 
\Big] \Big\}^{-1/2}
\end{eqnarray}
\begin{equation}
\mbox{with} \qquad 
\chi = \frac{(\sigma_l/\sigma_0)^2 - 1}{(\sigma_l/\sigma_0)^2 + 1}.
\end{equation}
The particles have the mass $m$ and the moment of inertia
$I = 50 \; m \sigma_0^2$. For convenience, we choose the units 
such that $\sigma_0 = \epsilon_0 = m = 1$. This defines the 
time unit $\tau = \sigma_0 \sqrt{m/\epsilon_0}$. The temperature
was $k_B T = \epsilon_0$.

In our previous work~\cite{akino_01a}, we have performed
an equilibrium molecular dynamics simulation of a system 
with 115200 particles in an elongated box with side ratios 
(1:2:1) and periodic boundary conditions, at a density 
in the coexistence region, $\rho = 0.017/\sigma_0^3$.
The system thus contained an isotropic and a nematic slab,
separated by two interfaces. The director in the nematic phase 
was found to align parallel to the interface (planar). 
Within that plane, the direction of the director is of course 
arbitrary, but the initial equilibration stage was conducted 
such that it pointed roughly in the direction of one box side.

The final configurations from these simulations were used as 
starting configurations in the present work.
Shear flow was enforced by applying Lees-Edwards boundary 
conditions~\cite{lees_72a,allen_book,evans_book}, \ie the periodic
boundaries in one direction move at constant speed. This
minimizes surface effects, but at the expense of constantly
pumping energy into the system. In order to dissipate
the latter, we have coupled the molecular velocities to a
Nos\'e-Hoover thermostat~\cite{frenkel_book}, taking care
that it acts only on the thermal fluctuations of the
velocities, \ie the velocities minus the streaming velocity 
of the fluid. The streaming velocity profile was determined 
from the simulation. In a nonuniform system like ours, using 
such a profile unbiased thermostat~\cite{evans_book} is
obviously important. However, we have noted in preliminary 
studies that we get practically the same results with a 
profile biased thermostat, where the actual velocity 
profile is approximated by a constant gradient profile. 
Hence the results do not seem to depend sensitively on 
the details of the thermostat.

\begin{figure}[t]
\begin{center}
\includegraphics[width=.35\textwidth]{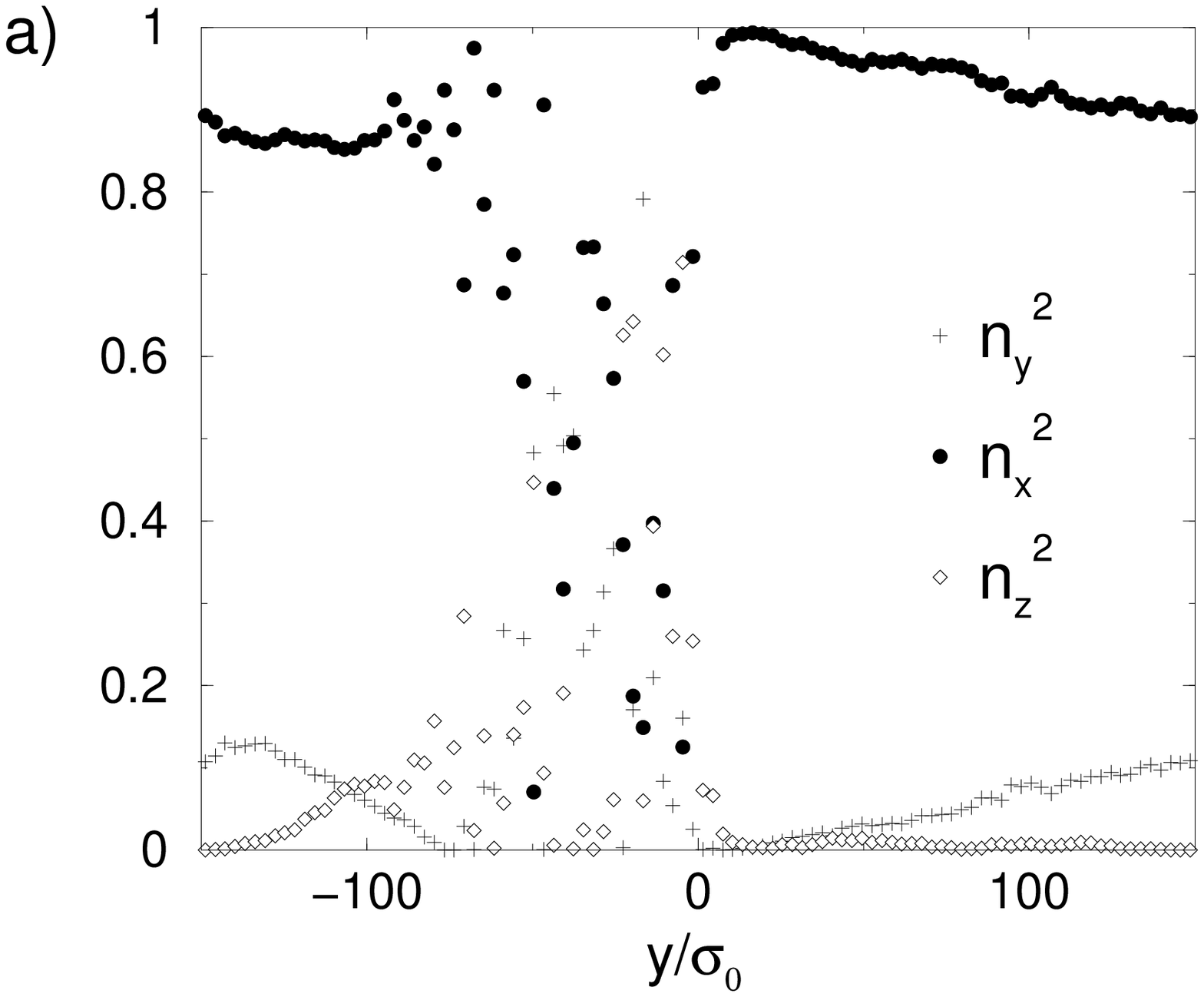}
\includegraphics[width=.35\textwidth]{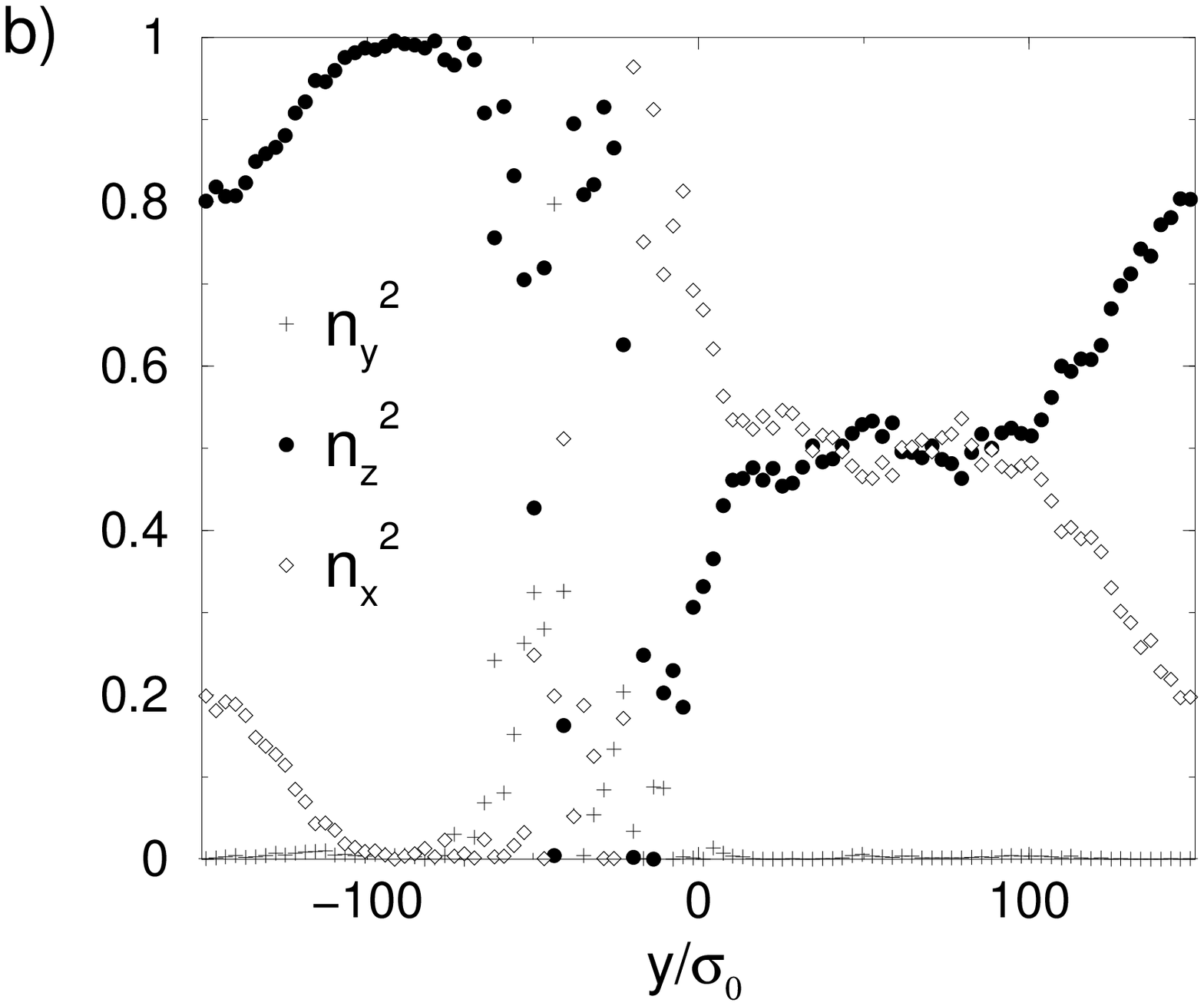}
\caption{
Director orientation profiles in the final configuration
of the interface set up in the (a) flow-aligned and
(b) log-rolling geometry, with $N=115200$ particles.
In the isotropic slab,
which is situated between $y=-80\sigma_0$ and $y=0$,
the direction of the director is not well-defined. 
The symbols show the squared component $n_\alpha^2$ 
of the director in different directions:
the shear gradient direction (plus, $\alpha=y$), the director 
direction in the initial setup (closed circles),
and the remaining direction (open diamonds).
After the initial ``equilibration'' time, these 
profiles stay roughly constant throughout the simulation.
}
\label{fig:dirprof}
\end{center}
\end{figure}

In the present paper, we discuss interfaces in the common 
stress geometry (Fig.~\ref{fig:geometries}, left). 
Two situations were considered: In the first, termed
``flow-aligned'' case, the initial configuration was turned
such that the director in the nematic phase points
in the direction of the flow, \ie the $x$ direction in
Fig.~\ref{fig:geometries}. In the second, ``log-rolling'' 
case, the initial configuration was turned such that
the director points in the direction of the vorticity, 
\ie the $z$ direction. We have first examined the stability 
of the interface in the flow aligned case for global strain 
rates up to $\dot{\gamma} = \Delta v_x/L_y = 0.1 /\tau$.
($\Delta v_x$ is the streaming velocity difference 
between both sides of the box, as imposed by the Lees-Edwards 
boundary conditions, and $L_y$ the box length in the $y$ 
direction.) To this end, we have carried out a 
series of simulations of smaller systems with 7200 particles
in a box with side ratios (1:8:1). Then we have focussed
on the strain rate $\dot{\gamma} = 0.001 / \tau$, and
performed extensive simulations of interfaces with initial
flow-aligned and log-rolling setup. In both cases
the system was first ``equilibrated'' over 2 million MD 
steps, and data were then collected over the next 
2 million MD steps, where one step corresponds to 0.0015
time units $\tau$. In the flow-aligned case, the system
remained flow-aligned throughout the simulation.
In the log-rolling case, two things happened during 
the initial ``equilibration'' time: The director of
the nematic phase turned very slowly into the flow 
direction, and the density of the nematic phase
decreased slightly (implying that the thickness of the 
paranematic slab shrank). After the initial ``equilibration''
time, the system had reached a state where half of the nematic 
slab was oriented at 45 degrees to the flow, and the 
director of the other half was still perpendicular to the 
flow (see Fig.~\ref{fig:dirprof}). 
This state did not change any more over the next 2 
million MD steps, in which the data were collected. 
Thus we could analyze the properties of this long-lived, 
yet presumably metastable state and compare them 
with the properties of the stable flow-aligned interface.

Since we use periodic boundary conditions, the 
interfaces can move freely through the system. 
Moreover, they exhibit capillary wave 
fluctuations~\cite{smoluchowski_08a,rowlinson_book,
akino_01a,elgeti_05a}, which broaden the apparent 
profiles and makes their width dependent on the
lateral system size. In order to eliminate these effects 
and determine local interfacial profiles, we adopt a block 
analysis technique introduced in our earlier studies
of equilibrium interfaces~\cite{akino_01a}:
We split the system of size $L \times L_y \times L$
into columns of block size $B \times L_y \times B$. 
In the $y$ direction, the columns are further
divided into 100 bins. Then we compute the local
order tensor~\cite{degennes_book} in each bin,
\begin{equation}
\Qtens = \frac{1}{N} \sum^N_{i = 1} \left( 
\frac{3}{2} \uu_i \otimes \uu_i - \frac{1}{2} \Itens \right),
\end{equation}
where $\Itens$ is the $3\times3$ unity matrix. 
The nematic order parameter $S(y)$ in the bin centered at $y$
is the highest eigenvalue of $\Qtens$ in that bin. 
From the profile $S(y)$, we determine the local positions
$y_{NI}$ and $y_{IN}$ of the two interfaces in the column 
under consideration, following the procedure described
in Reference ~\onlinecite{akino_01a}.
Then we calculate profiles for all quantities 
of interest and shift them by the amount $y_{NI}$ or
$y_{IN}$, respectively. This allows to perform averages
over {\em local} profiles. Here, profiles from
NI-interfaces are mirrored at $y_{NI}$ such that the
nematic slab is always on the right side.
The two interfaces from the log-rolling setup are discussed
separately, since they were different (cf. Fig.~\ref{fig:dirprof}).
The left interface, where the director on the nematic side is 
perpendicular to the flow, is labelled ``log rolling'', 
and the right interface, where the director is oriented
at 45 degrees with respect to the flow, is labelled ``45 degrees''.
In the flow-aligned case, we average over both interfaces.
The interface positions $y_{IN}$ and $y_{NI}$ themselves 
are used to analyze the capillary wave spectrum of the 
interfaces.

We should note that the two interfaces in the system are
slightly correlated, due to the finite width of the
isotropic and the nematic slab separating them. In particular,
the elastic interactions mediated through the nematic slab,
which already affect the capillary wave spectrum of a single
interface~\cite{elgeti_05a}, will also cause long-range 
interactions between the interfaces. A detailed finite-size 
analysis, and data with very low statistical error, would be 
necessary to investigate these effects in detail. 
Unfortunately, this would require such an immense computational
effort, that it is not feasible for us at the moment.

\section{Results}
\label{sec:results}

We first present our results for the interface at the strain rate 
$\dot{\gamma} = 0.001/\tau$. Some color snapshots of the system 
can be found in a recent preliminary report for the 
John-von-Neumann institute in J\"ulich, which is available
online~\cite{germano_04a}. 

The stress tensor, equal to the negative pressure tensor $\Ptens$,
can be calculated {\em via}
\begin{equation}
\sigtens = -\Ptens = -\frac{1}{V} \left( \sum^N_{i=1} m_i\mathbf{v}_i\otimes
\mathbf{v}_i + \sum^N_{i<j=1} \mathbf{r}_{ij}\otimes \mathbf{f}_{ij} \right).
\end{equation}
The shear stress is given by $(\sigma_{xy} + \sigma_{yx})/2$. 
For the flow-aligned setup, we obtained
$0.7 \pm 0.1 \cdot 10^{-4} \epsilon_0/\sigma_0^3$,
and for the log-rolling setup,
$1.3 \pm 0.1 \cdot 10^{-4} \epsilon_0/\sigma_0^3$.
The fact that the shear stress is higher in the log-rolling setup
({\em i.e.}, the average shear viscosity is higher) is another
indication that the log-rolling state is metastable. 
The antisymmetric stress $\sigma_{xy} - \sigma_{yx}$, as well
as all other nondiagonal elements of $\sigtens$, was zero
within the error. For comparison with the equilibrium interface,
we have also calculated the quantity
\begin{equation} 
\gamma = \frac{V}{2A} \big(P_{yy} - \frac{1}{2}(P_{xx}+P_{zz}) \big).
\end{equation}
In an equilibrium configuration with two interfaces in the $(xz)$ 
plane, $\gamma$ is the interfacial tension. The equilibrium 
value in our system was found to be~\cite{akino_01a} 
$\gamma = 0.009 \pm 0.003 \: \epsilon_0 \sigma_0$. 
Under shear, we obtain $\gamma = 0.003 \pm 0.001 \: \epsilon_0 \sigma_0$
in the flow-aligned setup, and 
$\gamma = 0.015 \pm 0.001 \: \epsilon_0 \sigma_0$ in the
log-rolling setup. The difference from the equilibrium 
value can be attributed to two effects: First, the interfacial properties 
change, and second, the diagonal components of the pressure 
in the bulk are no longer equal under shear. On principle, these
two contributions can be separated, either by systematically
varying the system size in the directions parallel and perpendicular
to the interface, or {\em via} a detailed analysis
of the profiles of $\gamma$~\cite{mcdonald_01a}. Unfortunately,
the first method is computationally too expensive, and the
statistics of the pressure profiles was not sufficient to
obtain results from the second method. Therefore, we have not 
been able to separate the different contributions to $\gamma$, 
and the interpretation of our result is difficult. 
The difference between the pressure tensor components
$P_{xx}-P_{zz}$ is $-3 \pm 1 \cdot 10^{-5} \epsilon_0 \sigma_0$
in the flow-aligned setup, and zero within the error in the
log-rolling setup.

\begin{figure}[t]
\begin{center}
\includegraphics[width=.4\textwidth]{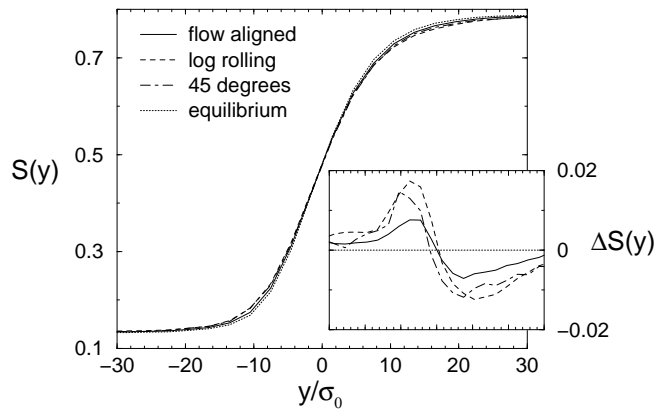}
\caption{Local order parameter profiles at 
average shear rate $\dot{\gamma} = 0.001/\tau$ 
in a system with $N=115200$ particles for the
flow aligned setup (solid), and the two
interfaces from the log rolling setup (dashed
and dot-dashed), compared to the corresponding 
equilibrium profile (dotted). Inset shows the 
difference $\Delta S$ between the 
profiles of the sheared and the equilibrium
interface. 
}
\label{fig:order}
\end{center}
\end{figure}

\begin{figure}[t]
\begin{center}
\includegraphics[width=.4\textwidth]{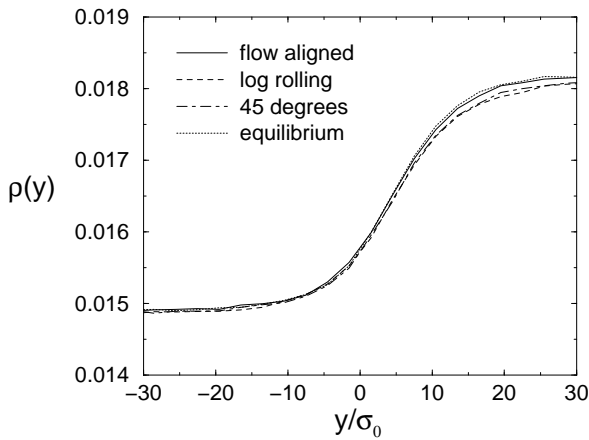}
\caption{Local density profiles 
at $\dot{\gamma} = 0.001/\tau$ ($N=115200$ particles). 
}
\label{fig:dens}
\end{center}
\end{figure}

The local interfacial profiles have been analyzed following
the procedure described in the previous section, with the lateral 
block size $B = 37.5 \sigma_0$ (\ie, the system was divided
into $4 \times 4 = 16$ columns). Figs.\ \ref{fig:order} 
and \ref{fig:dens} show profiles of the order parameter and 
the density at the interface for the flow-aligned and the 
log-rolling setup, and compare them with the corresponding equilibrium 
profiles. The order parameter profiles demonstrate that the 
interface broadens slightly under shear. In the log-rolling 
case, the effect is more pronounced than in the flow-aligned
case. Moreover, the density of the coexisting nematic phase
in the log-rolling case is slightly reduced.

\begin{figure}[t]
\begin{center}
\includegraphics[width=.4\textwidth]{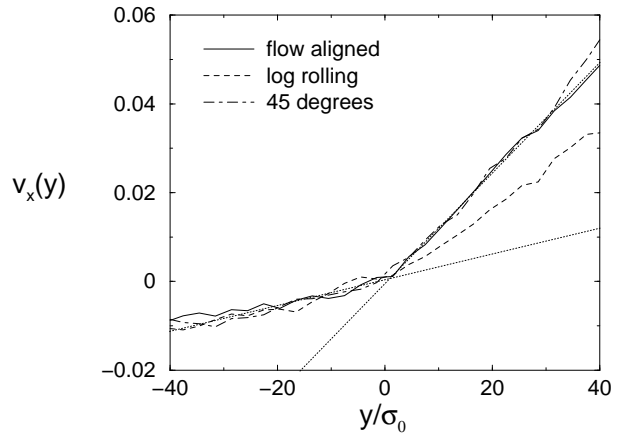}
\caption{Streaming velocity profiles in the flow
direction, shifted such that $v_x=0$ at $y=0$,
at $\dot{\gamma} = 0.001/\tau$ ($N=115200$ particles). 
The interface is located at $y = 0$. Profiles
from NI-interfaces are mirrored at $y=0$ and
$v_x=0$.
}
\label{fig:vx}
\end{center}
\end{figure}

\begin{figure}[t]
\begin{center}
\includegraphics[width=.4\textwidth]{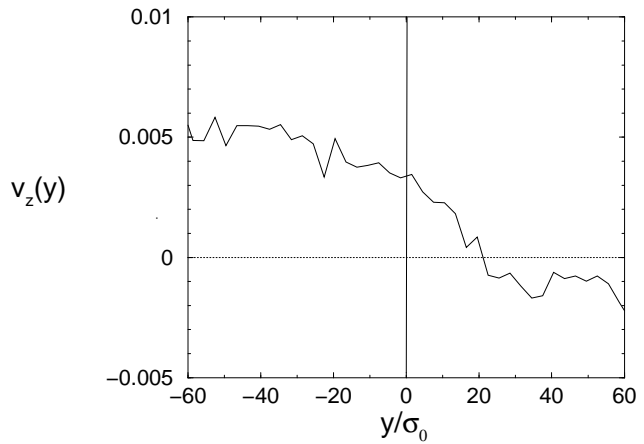}
\caption{Streaming velocity profile in the vorticity
direction at $\dot{\gamma} = 0.001/\tau$ for the
flow-aligned setup. 
}
\label{fig:vz}
\end{center}
\end{figure}

Next we examine the streaming velocity profiles at the interface,
shown in Figs.~\ref{fig:vx} and \ref{fig:vz}. 
The streaming velocity profile in the flow direction 
clearly exhibits two different slopes: The local shear 
rate is smaller in the paranematic phase than 
in the nematic phase. Thus we observe shear banding.

Even more remarkably, the shear also induces
flow in the {\em vorticity} direction $v_z$ 
(Fig.~\ref{fig:vz}). The direction of that flow
in the nematic and the paranematic phase is opposite. 
To our best knowledge, such a behavior has not been
reported for homogeneous systems. (Even if it had been 
observed, it could be removed by a simple Galilei 
transformation.) Hence our effect is clearly
induced by the interface. More precisely, the interface 
seems to induce a {\em flow gradient} in the vorticity 
direction. In the flow-aligned setup, where both interfaces 
are symmetrical, the velocity gradient dies off far from
the interfaces and the nematic and the isotropic slab move 
as blocks in opposite directions. In the log-rolling setup, 
the effect is even stronger (by a factor of three), 
but the situation is complicated by the fact that the two 
interfaces are different, and the integrated effect
of the flow gradient at the two interfaces is not
compatible with the periodic boundary conditions.
Therefore, we only show the data for the flow-aligned case here.

In the direction of shear gradient ($v_y$), no flow was
observed. Likewise, the streaming angular velocity $\vec{\Omega}$ 
was indistinguishable from zero within the error (0.0002). 
In fact, $\vec{\Omega}$ cannot be strictly zero in the isotropic 
phase, since general considerations show that it should be
one half the local shear rate~\cite{ailawadi_71a}.
In our system, however, the local shear rate in the 
isotropic slab was $\dot{\gamma} \approx 0.00025$ 
(cf. Fig.~\ref{fig:vz}), hence the resulting 
$\Omega = \dot{\gamma}/2$ is within the statistical error.

\begin{figure}[t]
\begin{center}
\includegraphics[width=.4\textwidth]{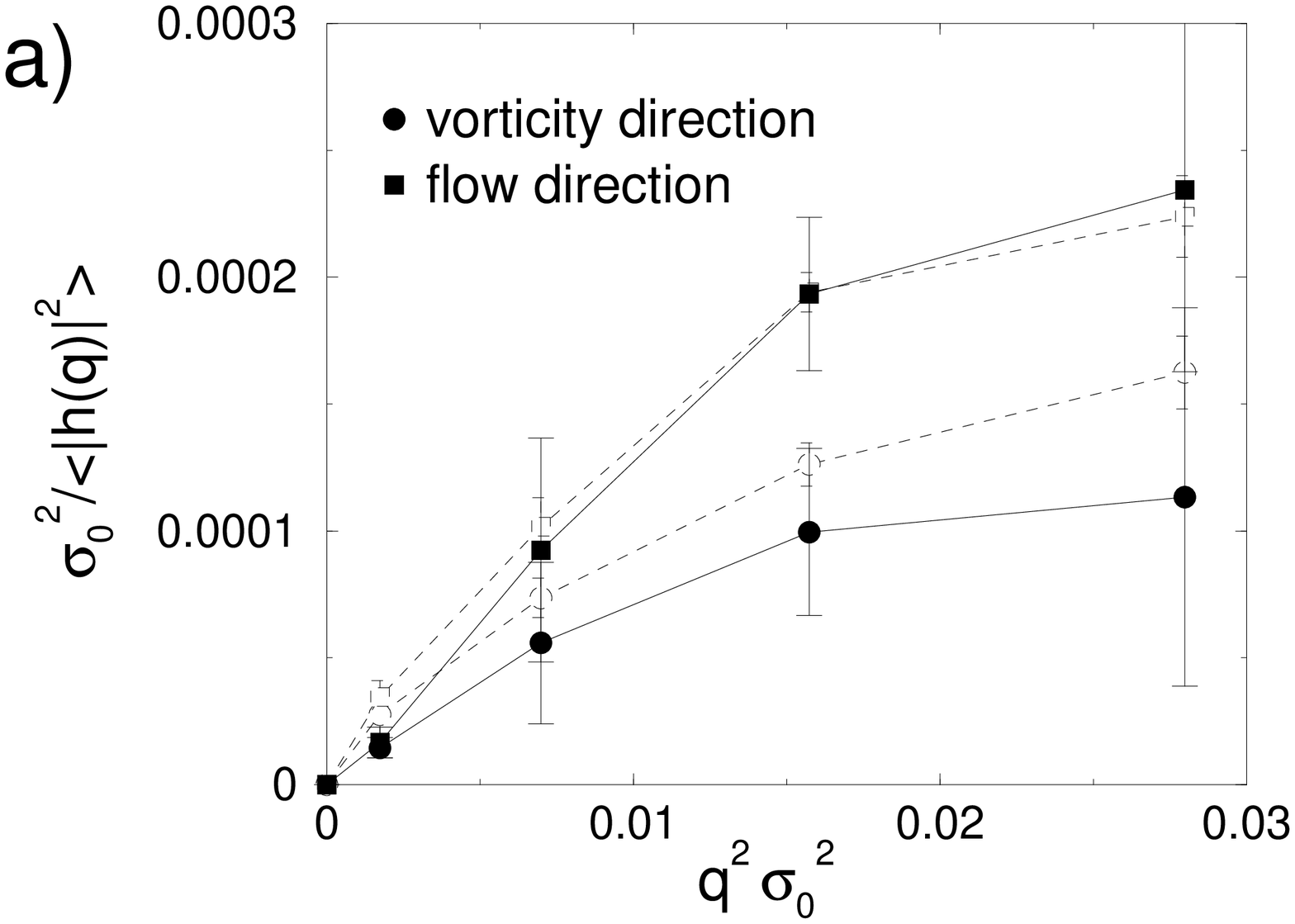}
\includegraphics[width=.4\textwidth]{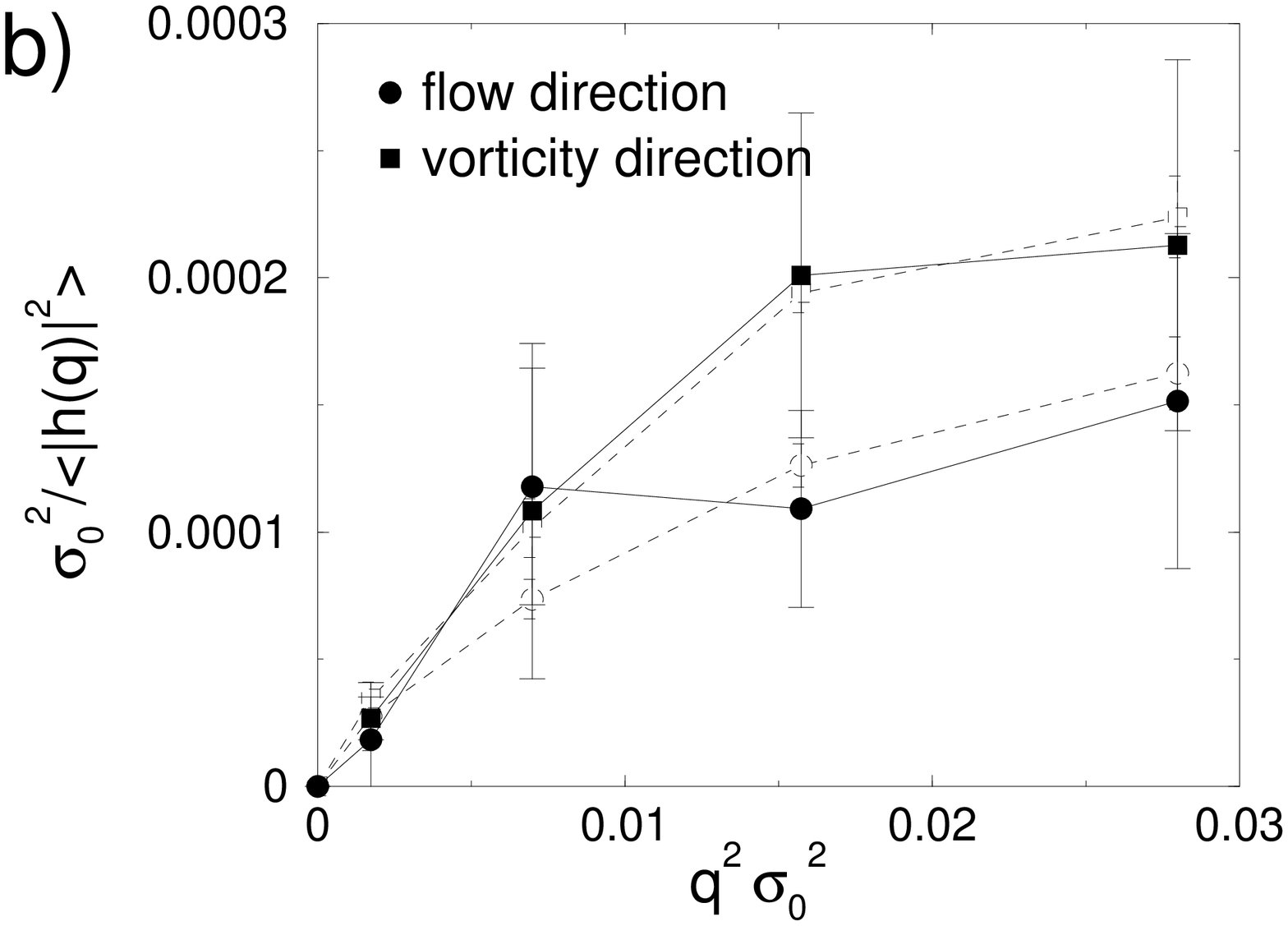}
\caption{
Capillary wave spectrum of the interface under shear 
with $\dot{\gamma} = 0.001/\tau$ ($N=115200$ particles) 
in the flow-aligned case (a) and the log-rolling case (b), 
for wave vectors pointing in the vorticity direction 
and in the flow direction. Dashed lines and symbols show 
for comparison the equilibrium spectrum  for
wave vectors parallel to the director
(squares, lower curve), and perpendicular
to the director (circles, upper curve).
}
\label{fig:hq2}
\end{center}
\end{figure}

The fluctuations of the interface positions can be 
characterized by the capillary wave spectrum: We assume
that we can parametrize the local position of an interface 
in the ($x,z$) plane by a single-valued function $y = h(x,z)$. 
The inverse of the fluctuations of its two dimensional Fourier 
transform, $\langle | h(\qq) |^2 \rangle^{-1}$, contains information 
on the local interfacial free energy. For example, an interface 
whose fluctuations are solely penalized by the interfacial 
tension $\gamma$ has the capillary wave spectrum~\cite{rowlinson_book}
\begin{equation}
\frac{k_B T}{\langle | h(\qq) |^2 \rangle} = \gamma q^2.
\end{equation}
The capillary waves of the equilibrium nematic/isotropic 
interface in our system have been investigated in detail 
in earlier work~\cite{akino_01a}. The spectrum
was found to be anisotropic, capillary waves being stronger 
in the direction perpendicular to the director than in that 
parallel to the director. This was recently explained 
theoretically by Elgeti and one of us within a 
Landau-de Gennes treatment~\cite{elgeti_05a}. According to
the theory, the spectrum should be isotropic in the 
long-wavelength limit $q \to 0$. However, this limit
is only reached on length scales of several thousand
correlation lengths. On smaller length scales, the
spectrum is dominated by strongly anisotropic cubic 
$q^3$-term and even higher order terms.

The capillary wave spectrum is the result of a subtle
interplay between the local director anchoring at the 
interface and the long-range elastic interactions.
Since the latter are affected by shear, we expected 
that the spectrum should change under shear.
We have performed a capillary wave analysis as described 
in the previous section with a lateral column size 
$B = 18.8 \sigma_0$ (\ie the system was split into 
$8 \times 8 = 64$ columns). The results are shown and 
compared with the spectrum of the equilibrium interface in 
Fig.~\ref{fig:hq2}. Somewhat surprisingly, shear does
not have much noticeable effect on the capillary wave
amplitudes. At low $q$-values, corresponding to large 
wave-lengths, the amplitudes $\langle | h(\qq)|^2 \rangle$ 
are higher than those of the equilibrium interface. 
This indicates that shear might reduce
the effective interface tension, and is compatible 
with the observation that the interfacial width 
broadens under shear. In the flow-aligned case, 
the capillary waves seem slightly enhanced in the 
vorticity direction. However, the effect is weak,
and the deviations from the equilibrium values
are still within the error bars.


\begin{figure}[t]
\begin{center}
\includegraphics[angle=-90,width=.45\textwidth]{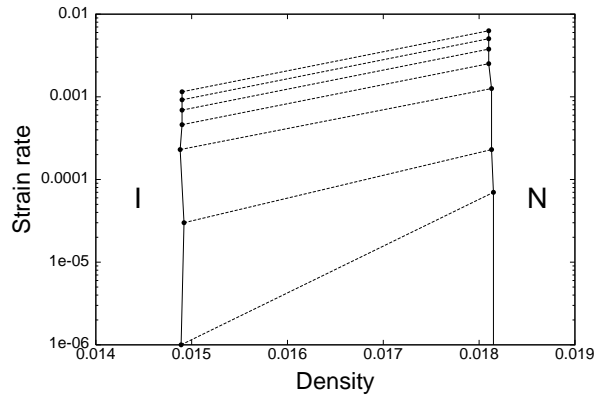}
\caption{
Phase diagram of the flow-aligned system. The dashed ``tie'' lines
connect coexisting states. I denotes the isotropic (paranematic)
region, and N the nematic region. The data were collected with
systems of size $N=7200$ particles.
}
\label{fig:ph_flal}
\end{center}
\end{figure}

Finally in this section, we discuss the phase coexistence
and the stability of the interface at other strain rates. 
To study this, we have carried out a number of simulations
of smaller, flow-aligned systems, where the length of the 
simulation box and the average density of particles was 
the same as before, but the lateral size was reduced by a 
factor of four (7200 particles in a box with aspect ratio 1:8:1). 
The interfaces remain stable up to an average strain rate 
of roughly $\dot{\gamma}=0.006/\tau$. 
Fig.~\ref{fig:ph_flal} shows the corresponding 
nonequilibrium phase diagram. Somewhat surprisingly, 
the densities of the two coexisting states are almost 
independent of the shear rate. This underlines our earlier
statement that the phase separation in our system is almost 
exclusively driven by thermodynamic forces. The local
shear rate is always higher in the nematic phase, \ie
the system is always shear thinning upon ordering.
In contrast to predictions from theoretical model 
calculations~\cite{olmsted_97a,olmsted_99a}, the 
coexistence region does not close up. 
Rather, the interface disappears abruptly beyond the 
average shear rate $\dot{\gamma}=0.006/\tau$. 
Coexistence regions similar to ours have been predicted for
coexisting paranematic and log-rolling states~\cite{olmsted_99a}. 
Why we observe this structure here for paranematic states
coexisting with flow-aligned nematic states is unclear,
and will have to be the subject of future studies.

\begin{figure}[t]
\begin{center}
\includegraphics[width=.5\textwidth]{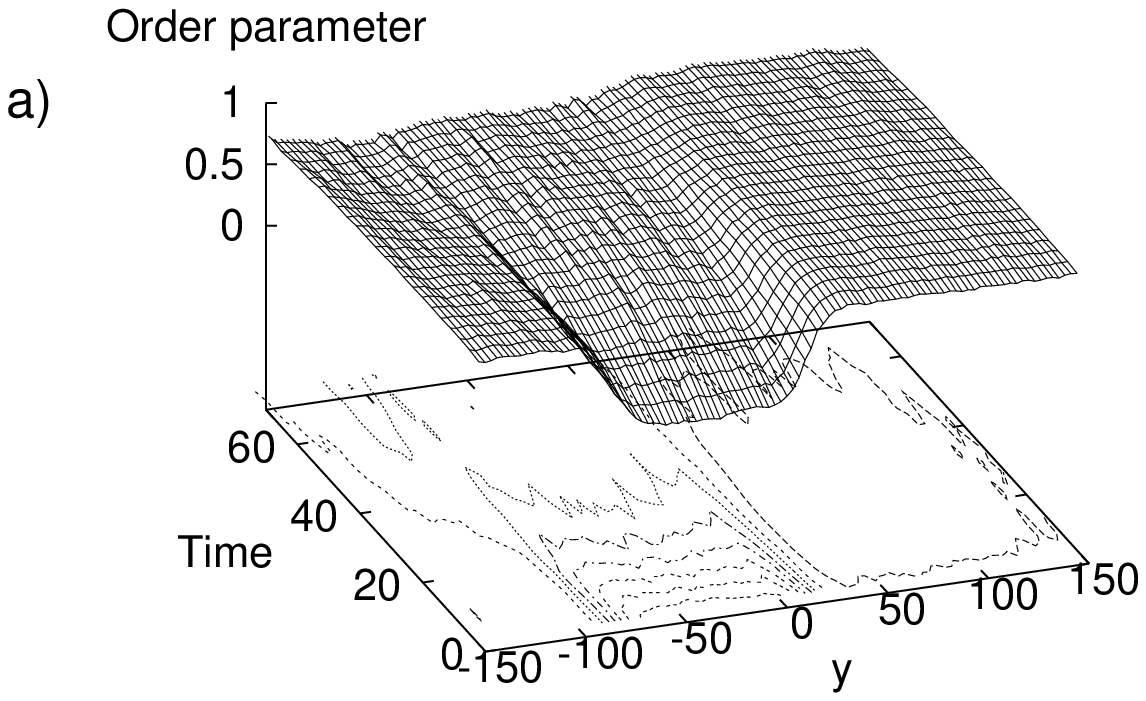}
\includegraphics[width=.5\textwidth]{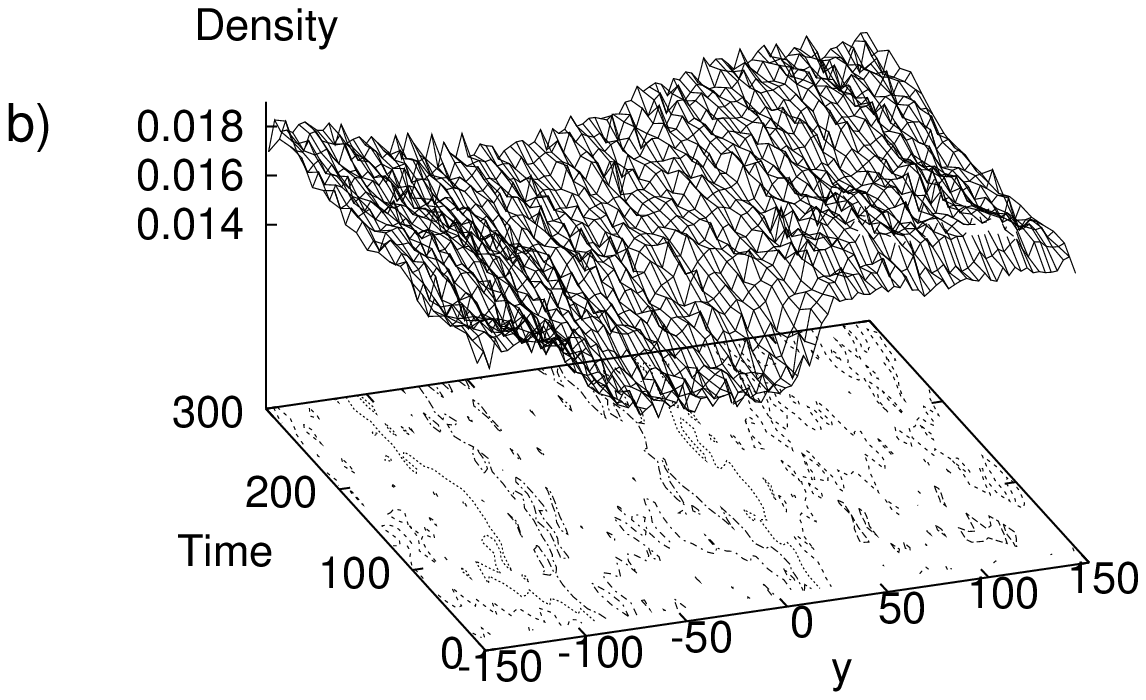}
\caption{
Destruction of the interface in the flow aligned system
at $\dot{\gamma} = 0.1/\tau$ ($N=115200$ particles).
a) Order parameter profile b) Density profile.
}
\label{fig:unstable_1}
\end{center}
\end{figure}

\begin{figure}[t]
\begin{center}
\includegraphics[width=.5\textwidth]{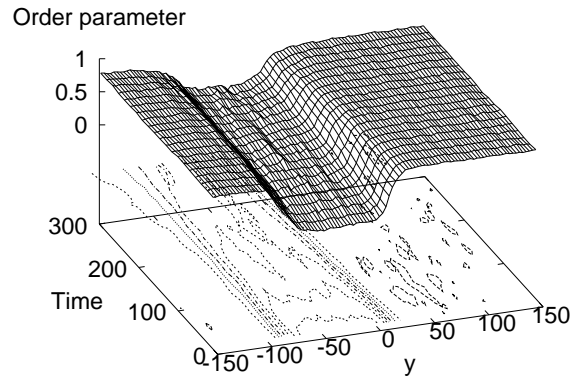}
\caption{
Destruction of the interface in the flow aligned system
at $\dot{\gamma} = 0.01/\tau$ ($N=115200$ particles).
}
\label{fig:unstable_2}
\end{center}
\end{figure}

Figs.~\ref{fig:unstable_1} and \ref{fig:unstable_2}
show the time evolution of interface destruction 
at shear rates $\dot{\gamma}=0.1/\tau$ and 
$\dot{\gamma}=0.01/\tau$. The initial configuration is
an equilibrium interface. At the shear rate
$\dot{\gamma} = 0.1/\tau$, the shear aligns the
liquid crystal throughout the paranematic slab, thereby
destroying the difference between the two phases. 
This is the primary process and happens on a time
scale of roughly $30 \tau$. In a second step,
the density slowly adjusts. The density profile therefore
persists much longer than the order parameter profile 
(Fig.~\ref{fig:unstable_1} b).
At the shear rate $\dot{\gamma} = 0.01 /\tau$,
the destruction process is less homogeneous
and nucleates at the interfaces: They
slowly move inwards and gradually lose their 
structure (Fig.~\ref{fig:unstable_2}). 
This happens on a time scale of roughly
$300 \tau$. The time scale of destruction at average
shear rate $0.01/\tau$ is thus about ten times as large 
as at shear rate $0.1/\tau$. We conclude that the time 
scale of destruction is roughly proportional to the
time scale $1/\dot{\gamma}$ introduced by the shear. 
This underlines the necessity of performing
very long simulation runs at lower shear rates.
In particular, it is important to note that at
the shear rate $\dot{\gamma} = 0.001 /\tau$, 
where we have performed most of our simulations,
the interface showed no sign of destabilization
after a run of length $6000 \tau$. Therefore we
can deduce that the interface is indeed stable.

\section{Conclusions and Outlook}
\label{sec:conclusions}

We have presented an extensive nonequilibrium 
molecular dynamics simulation of liquid crystal interfaces
in shear flow, and compared the properties 
of these interfaces with the structure of the corresponding 
equilibrium interface. Two situations were considered: 
In the first, flow-aligned setup, the interface was set up 
such that the director in the nematic phase pointed in the 
direction of shear gradient;
in the second, log-rolling setup, the director pointed in the 
direction of the vorticity. The log-rolling state
appeared to be unstable with respect to the flow-aligned
state, but extremely long-lived, so that it could be analyzed
as well. In both the flow-aligned and the log-rolling
system, the two coexisting states experience the same,
common, stress.  The third possible situation, phase coexistence
at common strain (Fig.~\ref{fig:geometries}, right),
was not yet studied. This will be done in future work.

We found that the structural properties of the interface
are not dramatically affected by the presence of shear. 
The interfacial width increases slightly, and the capillary
wave amplitudes are slightly enhanced at large wavelengths.
No other noticeable effects were observed. For example, the 
biaxiality was still basically zero (data not shown).

In contrast, the streaming velocity profiles revealed 
a number of interesting features: First, we observe
shear banding, \ie the shear profile is inhomogeneous. 
The local shear rate in the nematic phase is distinctly
higher than in the paranematic phase. 
Second and surprisingly, the interface also induces 
a streaming velocity gradient in the vorticity 
direction. As a result, vorticity flow in opposite 
directions is induced in the paranematic and the 
nematic phase. This flow is symmetry breaking, 
it destroys the mirror symmetry at the $(xy)$-plane. 
The effect is small but significant. So far, we have no 
explanation. It has not been predicted by theory; 
however, the theoretical studies of shear banding cited 
in the introduction were all set up such that the 
vorticity flow was zero by construction. 
We are not aware of any fundamental argument against
nonzero vorticity flow. Hence our result is perplexing, 
but not impossible. We hope that it will stimulate 
further theoretical work on shear banding, and that the 
mechanism and the conditions for symmetry breaking will 
be clarified.

The effect is not only of academic interest, it could
also be useful from a practical point of view. 
For example, if one component enriches in
one phase, and the other in the other phase, shear
could possibly be used to separate the two components. 

The present study has focused on interfaces which
are mainly stabilized by thermodynamic driving forces.
As mentioned in the introduction, nonequilibrium
interfaces can also be stabilized mechanically. 
In that case, the interfacial properties presumably
differ much more strongly from those of equilibrium
interfaces than in the present case. We hope that
we will succeed in stabilizing and studying
such an interface in the future.

\section*{Acknowledgments}

We thank Peter Olmsted and Michael Allen for useful discussions, 
and Radovan Bast for running many $N=7200$ systems within an 
undergraduate physical chemistry project. The parallel MD program
GBmega used in this work was  originally developed by the EPSRC
Complex Fluid Consortium. This work was funded by the German
Science Foundation, and the simulations were carried
out at the John-von-Neumann computing center in J\"ulich.

\end{document}